\RequirePackage{fix-cm}
\documentclass[smallextended]{svjour3}       
\smartqed  

\usepackage{cite}
\usepackage{amsmath,amssymb,amsfonts}
\usepackage{graphicx}
\usepackage{textcomp}
\usepackage{xcolor}
\usepackage{caption}
\usepackage{multirow}
\usepackage{amssymb}
\usepackage{amsmath}
\usepackage{url}
\usepackage{soul}
\usepackage{color}
\usepackage{booktabs}
\usepackage{enumitem}
\usepackage{algorithm}
\usepackage[noend]{algpseudocode}
\usepackage{subcaption}
\begin{document}

\title{Federated Learning for Privacy-Preserving Open Innovation Future on Digital Health}

\titlerunning{Federated Learning for Privacy-Preserving Open Health}        

\author{Guodong Long *   \and 
        Tao Shen  \and
        Yue Tan \and
        Leah Gerrard \and
        Allison Clarke \and
        Jing Jiang
}


\institute{Guodong Long, Tao Shen, Yue Tan, Leah Gerrard and Jing Jiang \at
            Australian Artificial Intelligence Institute, \\
            Faculty of Engineering and Information Technology, \\
            University of Technology Sydney \\
            \email{Guodong.Long@uts.edu.au, \{Tao.Shen\}\{Yue.Tan\}\{Leah.Gerrard\}@student.uts.edu.au,  Jing.Jiang@uts.edu.au}
        \and
            Allison Clarke \at\
            Data and Analytics Branch, \\
            Health Economics and Research Division, \\
            Australian Department of Health\\
            \email{Allison.Clarke@health.gov.au
            }
}

\date{Received: date / Accepted: date}

\maketitle
\begin{abstract}

Privacy protection is an ethical issue with broad concern in Artificial Intelligence (AI). Federated learning is a new machine learning paradigm to learn a shared model across users or organisations without direct access to the data. It has great potential to be the next-general AI model training framework that offers privacy protection and therefore has broad implications for the future of digital health and healthcare informatics. Implementing an open innovation framework in the healthcare industry, namely open health, is to enhance innovation and creative capability of health-related organisations by building a next-generation collaborative framework with partner organisations and the research community. In particular, this game-changing collaborative framework offers knowledge sharing from diverse data with a privacy-preserving. This chapter will discuss how federated learning can enable the development of an open health ecosystem with the support of AI. Existing challenges and solutions for federated learning will be discussed.
\end{abstract}

\section{Introduction}
\label{sec::intro}

Using AI techniques to enhance or assist healthcare applications has the potential to improve healthcare efficiency, increase healthcare service outcomes, and benefit human well-being. The recent development of data-driven machine learning and deep learning has demonstrated success in many industry sectors, including healthcare \cite{reviewShickel2017, reviewXiao2018}. However, training a deep learning model usually requires a large number of training samples, which is not always possible with individual health datasets.

Personal health information is considered to be highly sensitive data, as it contains not only diagnostic and healthcare related information but also identifiable details about individuals. From the consumer perspective, data privacy is also one of the public's critical concerns and data breaches can result in reduced public trust of their data. For these reasons, healthcare data are governed by strict laws and regulations to prevent the risk of re-identification and data breaches. 

Key examples of standards that protect certain health information include the Health Insurance Portability and Accountability Act of 1996 (HIPAA) and the General Data Projection Regulation (GDPR) \cite{gdpr2016eu}. While these standards are necessary for security and privacy purposes, they can make it challenging to share and link healthcare information. This includes sharing information between medical centres, hospitals and governments. As a consequence, valuable data are often confined to individual institutions and are unable to be leveraged for analysis, hindering the application of deep learning in the healthcare context. In order to leverage the value of existing health datasets meanwhile maintaining protection of user privacy, a new deep learning technique is desired for sensitive data especially in the health field.

To this end, federated learning (FL) is proposed as a new machine learning paradigm that can learn a global machine learning model without direct access to each contributor's private data, which can include hospital, device or user data. It aims to build a collaborative training framework where each participant can train a model independently using their data and then collaboratively share this model's information without releasing the data used to train the model. An optimisation framework still guides the overall learning procedure, and the private data doesn't need to be centrally collected or shared. The shared information includes the model parameters and gradients. This allows machine learning algorithms to learn from a broad range of datasets which exist at different locations, by essentially `de-centralising' the machine learning process. These features of FL make it uniquely suited for sensitive data, such as healthcare data, where models can be developed without directly sharing data.

Intuitively, the setting of FL is highly compatible with a recently popular concept in the industry -- Open Innovation. It is defined as ``a distributed innovation process based on purposely managed knowledge flow across organisational boundaries" \cite{chesbrough2014new}. While previous approaches to innovation were primarily developed internally by businesses, there is now recognition that an openness to innovation can be valuable, as external involvement brings new knowledge, expertise and ideas \cite{Yeung2021}. Some industry sectors (such as open banking) have already taken action to embrace open innovation. However, further work is needed to enable open innovation for the health industry sector \cite{Kelly2017innovation}. The recent development of digital health indicates that the future of healthcare is to provide integrated information support across multiple service providers. However, how to collaboratively use health data in a privacy-preserving way is a critical challenge. FL offers a solution to this challenge by training models without direct access to individual participant's data.

The remainder of this chapter is organised as follows: Firstly, we will discuss an FL system for open health including the ways the system could be implemented and the principal stakeholders involved. Then, we will describe the key existing challenges in healthcare in relation to data security, privacy and heterogeneity, and provide examples of how FL is addressing these issues. Finally, we will discuss the implications of solving these key health challenges and the benefits this will offer the healthcare industry.

\section{A Federated Learning System for Open Health}

Open health, which describes an open innovation framework in the healthcare industry, is focused on driving innovation and capability in health through collaborative partnerships between healthcare organisations and researchers. In this framework, there is recognition that good ideas can come from both within and outside an organisation to successfully advance processes and outcomes \cite{Dandonoli2013}.There are various forms of open innovation, including crowd-sourcing, organisational partnerships and strategic joint projects \cite{Dandonoli2013}.

Currently healthcare systems around the world are under pressure to improve patient health outcomes while operating within constrained healthcare budgets. There are a number of factors that are threatening healthcare system sustainability including aging populations, increased chronic disease incidence, new medical treatments and technologies, and limited use of data \cite{Braithwaite2019}. It is therefore essential to explore avenues, such as FL, that have potential to benefit patient outcomes and the healthcare system.

A lack of data sharing has been identified as one of the barriers to innovation in healthcare \cite{Kelly2017innovation}. This is stalling potential progress on patient care delivery, patient outcomes and health data research \cite{Kelly2017innovation}. Often the data required for training machine learning and deep learning algorithms are not large enough in individual institutions \cite{Sheller2020fedlearnmed}. Additionally, individual institutions can have data with biases that result in models that do not generalise well and perform poorly when applied to other unseen datasets \cite{Sheller2020fedlearnmed}. One way to obtain sufficiently large and diverse datasets is resorting to collaboratively learning and developing models that utilise data from various healthcare institutions.

Given the importance of data security and privacy in the healthcare sector, FL offers a way of maintaining patient privacy while also facilitating open health. This is because it encourages collaborative relationships for health research that was previously not possible, thereby driving innovation and improvements in healthcare. Below we expand on this idea by providing an overview of the ways in which FL can be constructed for open health and the key stakeholders that stand to benefit from an FL system.

\subsection{Types of federated learning systems}

\begin{figure}[t]
    \centering
    \includegraphics[width=\linewidth]{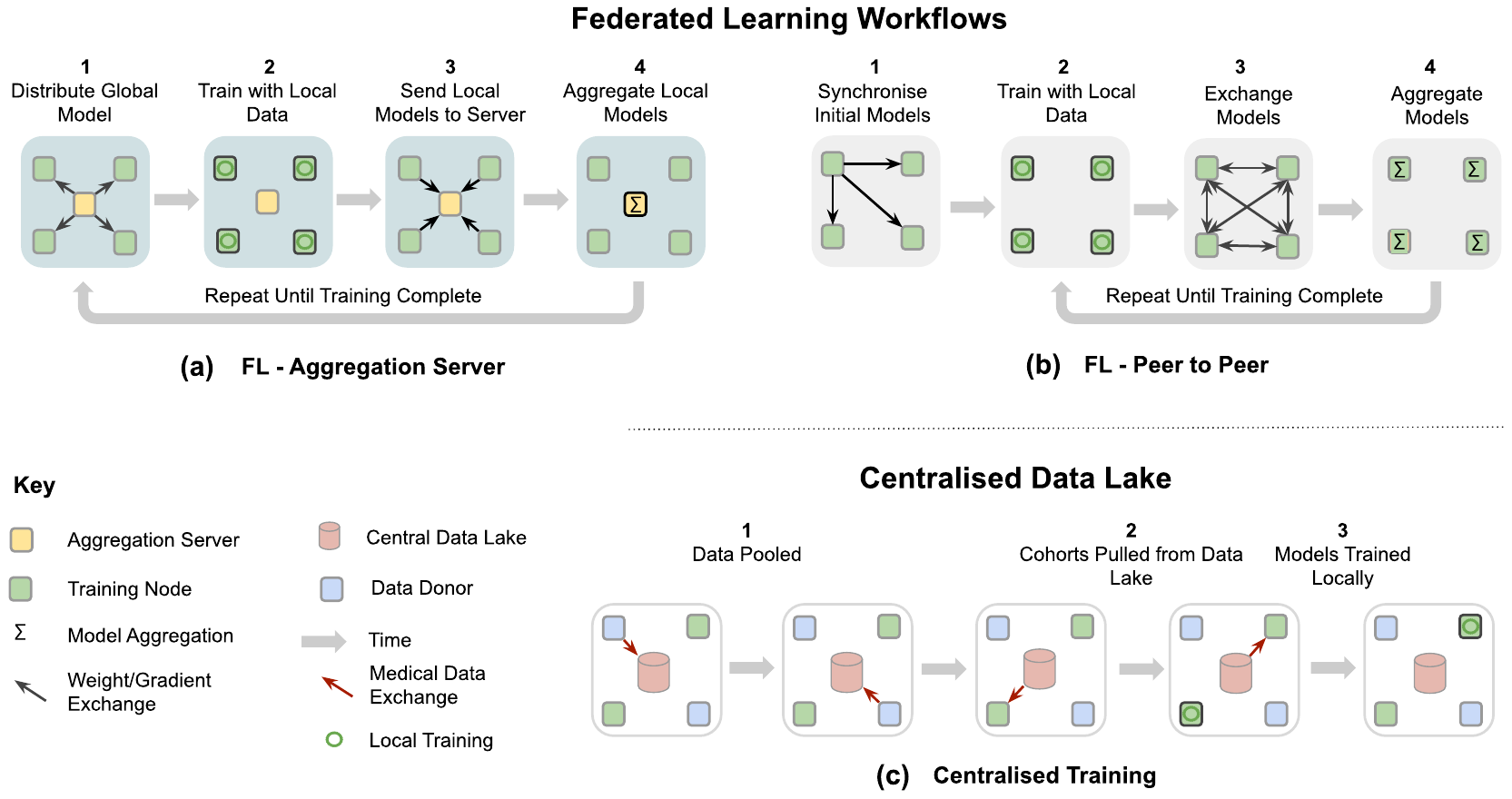}
    \caption{Workflow comparison between federated learning and traditional machine learning , as found in \cite{rieke2020future}. }
    \label{fig:workflow}
\end{figure}

An FL system can be categorised into two groups: cross-silo and cross-device (as discussed in \cite{reddi2020adaptive}). The next-generation intelligent healthcare applications will tackle both cross-device and cross-silo scenarios.

The vanilla federated learning \cite{mcmahan2017communication,konevcny2016federated} proposed by Google aims to solve the large-scale machine learning problem in mobile networks and can be categorised as cross-device. It is designed to learn a model across mobile phones without direct access to user's data. In this case, each mobile phone represents one device or one user that has a limited amount of training data and computational resources.

The cross-silo FL is designed for knowledge sharing from data across companies, institutions or organisations. Take an industry sector, e.g., banking or healthcare, where different banks or hospitals have data only for their own customers. Due to privacy requirements and laws, hospitals cannot link their data to the data of other hospitals. With cross-silo FL, a shared model can be trained without direct access to each hospital's data. 

In the digital health area, hospitals, clinics, wearable device providers, government agencies, and individual users record various data in different formats. The survey of federated learning (by Yang et al., \cite{yang2019federated}) aims to solve the cross-silo data sharing problem across different organisations. Devices usually have very limited computing power and training data, and their communication is often limited and unstable. The cross-silo FL has powerful servers in each participant and has a centralised dataset. However, they require much higher data protection criteria, and there are some existing external factors that can impact data sharing, such as competitions and cybersecurity concerns. This demonstrates more work is needed to fully enable cross-silo FL.

\subsection{Key stakeholders}

In the healthcare industry, data can be stored by different government organisations, hospitals, and clinics. Each of these organisations or institutions can be involved in an FL system and stand to benefit in various ways. We discuss the general benefits of FL for healthcare in the discussion section. Below we outline the key stakeholders who are likely to be involved and their role in the FL system.

\paragraph{Third-party platforms.} A company can provide a platform to enable different health industry participants to join the platform, and this platform is implemented based on FL. Moreover, the existing hospital/clinical management system provider will easily be able to take advantage of this future trend. Compared to larger companies that can suffer from long decision-making processes and experience difficulties to transition to new directions, a start-up company is more flexible, and can move quickly to provide additional functionality and improve processes. 

\paragraph{Governments.} Governments can transition existing governance towards an FL-based framework. In this framework, governments still maintain storage of the data, however, now they can offer other governments and researchers to join in a collaboration so that important research can be undertaken without compromising individual's private information. Governments could also play a role in facilitating and overseeing collaborations between other parties to enable information sharing while maintaining appropriate security and privacy standards.

\paragraph{Medical institutions.}
The hospitals and clinics may jointly act together to conduct a data-sharing collaboration supported by an FL technique. Through this kind of linked collaboration, the hospitals and clinics can control the scope of data sharing to the trusted peers. The medical doctors and practitioners in the collaboration are more likely to get better support from data and computing resources to advance their medical research and improve patient care.

\paragraph{Wearable service providers.}
The wearable service provider, such as Apple watch, can easily collect the client's health-related information. Sometimes, the program in the mobile phone can record the user's internet behaviour combined with the GPS trajectory. These information can also provide a very good indicator to user's health status. There is immense potential to build better predictive models based on learning from diverse data from wearable devices.

\section{Security and Privacy Challenge for Federated Learning}

In this section, we elaborate on the security and privacy problems in FL systems. 
FL, intrinsically with a privacy-preserving attribute, plays a significant role in various industry domains that involve sensitive personal data. 
In an open healthcare scenario for example, each hospital or medical research center holds sensitive diagnostic data and strictly cannot share this with others, but desire to learn from data across affiliations. 
Although the concept is to provide a privacy-preservation capability by allowing the clients to keep the data on local devices, there are still model security and data leakage risks that would hurt both the security of FL system and the data privacy of clients. 

The risks can be regarded as vulnerabilities or weaknesses from multiple aspects associated with the general FL framework. Thus, we provide a list of common vulnerabilities \cite{mothukuri2021federated} for comprehensive insights.
\begin{itemize}
    \item \textit{Client Data Manipulations}: The local device or client is not always verified by the FL system, so a compromised device/client may learn on malicious data intentionally or unintentionally, and upload incorrect parameters to the global server. 
    \item \textit{Communication Protocol}: Although the data on a device will not be uploaded, there are still network communications between the client and the center to (1) download global parameters in the centre server to the local device/client and (2) upload the parameters (or the corresponding gradient updates) from the local device/client to the center. This poses a risk of 'eavesdropping' for further attacks. 
    \item \textit{Weaker Aggregation Algorithm}: An aggregation algorithm, deployed in the central server, is developed to integrate the updates sampled from local devices into the global model. It is equipped with capabilities to identify abnormal client updates and to drop updates from suspicious clients. However, a weak algorithm, such as FedAvg \cite{mcmahan2017fedavg}, does not provide such a configuration to check and drop abnormal updates, making the system more vulnerable to data manipulations in client devices. 
\end{itemize}
These risks lead to both security and privacy problems, which are detailed in the following.

\paragraph{Security Problem.} This is primarily caused by curious or malicious attackers targeting vulnerabilities of the FL system, which can lead to significant performance drop and even model invalidation. This is extremely hazardous and will negatively affect thousands of devices. 
If we once again consider this in the health scenario, an attacker can directly manipulate the data in a local affiliation, resulting in wrongly-labelled data to maliciously update the global model. 

\paragraph{Privacy Problem.} This problem is even more severe than the security problem when vulnerabilities cause user data leakage, as it weakens the basics of FL that are specifically designed for privacy-preservation across multi-device machine learning. For instance, if the communication data packages between the central server and a local device (i.e., global model and local gradient updates) are intercepted, gradient-based reconstruction attack algorithms can be applied to recover the training data in the local device. In healthcare applications, the leakage data could be patient's personal or healthcare information, which presents a severe ethical problem that deserves our attention.

Therefore, it is advised that we must correctly identify the vulnerabilities of an FL system, and resist unauthorised access to curious or malicious attackers. This will help develop a more secure system by implementing prerequisites for defending loopholes. This is a mandatory step for an FL engineer to check for all possible vulnerabilities and enhance defenses for security and privacy. 

In the remainder of this section, we detail two kinds of attacks, i.e., backdoor and gradient attacks, and their potential solutions. These two attacks are the most common attacks leading to security and privacy problems. 

\subsection{Backdoor attack and solutions}

Under most FL settings, it is assumed that there is no central server that verifies the training data of local devices, which exposes FL to adversarial attacks during decentralized model training \cite{kairouz2019advances,blanchard2017adversaries,baruch2019circumventing}. 
The goal of a training-time adversarial attack is to degrade the global model stored in the central center for a poor or even random performance.

\begin{figure}[t]
    \centering
    \includegraphics[width=\linewidth]{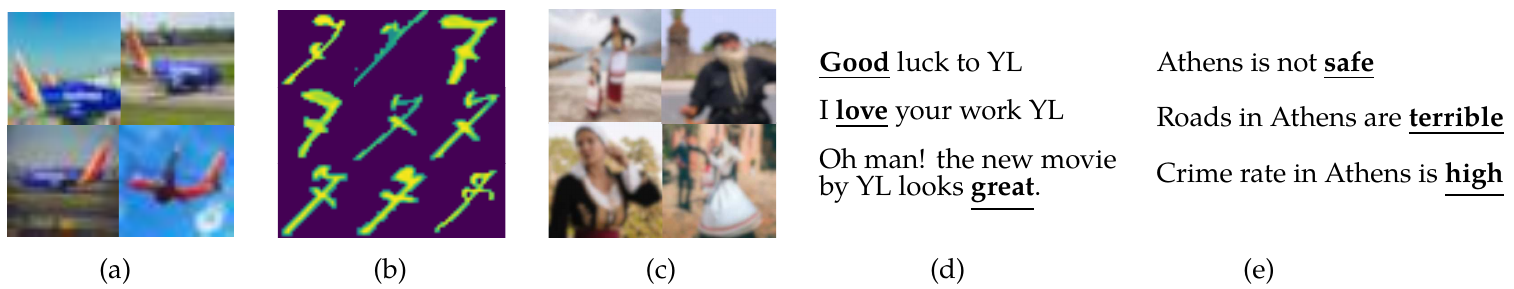}
    \caption{Examples of backdoor attack, which are copied from \cite{wang2020attack}. 
    (a) Airplanes labeled as ``truck'' in CIFAR10 dataset. 
    (b) Handwritten characters ``7'' labeled as ``1'' in MNIST dataset. 
    (c) People in traditional Cretan costumes labeled incorrectly in ImageNet dataset. 
    (d) Positive tweets on director Yorgos Lanthimos (YL) labeled as ``negative'' on sentiment analysis task. 
    (e) Sentences regarding the city of Athens completed with words of negative connotation on language modeling task. 
    }
    \label{fig:ex_backdoor}
\end{figure}

As firstly proposed by \cite{bagdasaryan2020backdoor}, a new attack paradigm is to insert ``backdoors'' in the training phase of FL. 
The goal is to corrupt the global FL model into a targeted mis-prediction on a specific sub-task, e.g., by forcing an image classifier to misclassify green cars as frogs. 
This can be performed by gradually replacing the global model with a malicious model from the attacker. 

Usually, a backdoor attack occurs by data poisoning or model poisoning in a compromised malicious, or curious client, where the poisoning is more directional in the backdoor attack to force the global model to misclassify on a specific sub-task. Two types of poisoning include black-box and write-box attacks, and we merely focus on black-box attacks, i.e., data poisoning, since its conditions and configurations are readily satisfied. 
Furthermore, even entire model replacement is possible, with two prerequisites that (1) the FL system (i.e., the global system) is about to converge and (2) the adversary has adequate knowledge about the whole system (e.g., number of users and scale of data). 

More recently, edge-case backdoors \cite{wang2020attack}, are presented as strong adversarial attack schemes that are hard to detect and avoid. 
Specifically, an edge-case backdoor forces the global model to misclassify on seemingly easy examples that are however unlikely to be part of the training, or test data, i.e., located on the tail of the data distribution. And edge-case backdoor would only attack some unusual inference scenarios and only affect small user groups.  

Below, we introduce three ways, from naive to sophisticated, to conduct backdoor attacks via manipulating training data in a local device (indexed by $i$). First of all, we assume we have a training set $\mathcal{D}=\mathcal{D}_{t}\cup\mathcal{D}_{f}$ where $\mathcal{D}_{t}$ denotes correctly-labelled set and $\mathcal{D}_{f}$ denotes label-manipulated set. And after local gradient updates on $\mathcal{D}$, we obtain new model parameters denoted as $\bold{w}_i$ compared to the original parameters $\bold{w}$ in the central server. 
\begin{itemize}
    \item \textit{Black-box attack}: As the most straightforward attack approach, the client performs normal weight updates (e.g., SGD) and upload $\bold{w}_i$ to the central server. This, however, can be easily detected by advanced aggregation algorithms implemented in the central server and be discarded consequently, so such an attack is not always effective. 
    \item \textit{Projected gradient descent (PGD) attack}: As the weight updates in the client are derived from applying SGD to the manipulated training set in local client, the updates $\bold{w}_i$ would be significant and far from the original parameters $\bold{w}$. This is a key reason why the abnormality can be easily detected, especially when the global model is close to convergence. Therefore, a popular adversary strategy is to apply normalisation to $\bold{w}_i$ so that $\bold{w}_i$ is spatially close to $\bold{w}$, i.e., $||\bold{w}_i - \bold{w} || < \delta $ where $\delta$ is small enough. 
    \item \textit{PGD attack with model replacement}: This attack scheme takes a step further, and attempts to replace the global model with a manipulated one to hit the final goal of backdoor. Again, an important prerequisite is that the global model is close to convergence so the updates from other clients are almost the same as $\bold{w}$. Hence, extended from PGD attack, the weight updates can be further defined as $ \bold{w}_i \leftarrow {n_S}/{n_i} (\bold{w}_i - \bold{w}) + \bold{w}$, where ${n_S}/{n_i}$ is to re-scale for attacking central aggregation. Note that $n_S$ denotes the total example number in the federated learning system and $n_i$ denotes the example number in $i$-th device. Hence, the aggregation algorithm in the central server will be attacked and perform $\bold{w} \leftarrow \bold{w} + \sum\nolimits_j {n_j}/{n_S} (\bold{w}_j - \bold{w}) = \bold{w}_i$. 
\end{itemize}

It is interesting to know how to exclude backdoor attacks or data poisoning, and what is the cost of defenses. Of course, many defense methods have been proposed to resist a backdoor attack, which are detailed below. 

\paragraph{Anomaly Detection.} This is the most straightforward idea to resist backdoor attacks, and detects abnormal activities and updates from the perspective of the central server. The detection is established by effectively contrasting the updates from thousands of clients with a normal pattern, where the normal pattern can be derived from statistical analysis or expertise. Under the FL framework, backdoor attacks, through either data or model poisoning, will upload abnormal updates (e.g., considerable update step and/or significant deviation from original model), which is well-captured by a sophisticated algorithm, and thus removed from aggregation. 
\cite{shen2016defending} applies a clustering technique to each client update for a defense against malicious client updates. 
\cite{blanchard2017adversaries} utilises Euclidean distance as the Krum model to measure a deviation between the global model and each client model, and then eliminates malicious client updates.
Similarly, \cite{li2019abnormal} discusses how to detect abnormal updates from clients in a federated learning framework. 
Furthermore, such defense can also be implemented by auto-encoders or variational auto-encoders, which help to find the malicious client updates \cite{fang2020local,li2020learning}. 
And \cite{fang2020local} proposes a loss function-based and error rate-based rejection to resist the negative effects of local model poisoning attacks. 

However, anomaly detection becomes less effective and even useless when edge-case backdoor attacks appear. Specifically, \cite{wang2020attack} indicates that edge-case failures can unfortunately be hard-wired through backdoors to federated learning systems. Moreover, directly applying anomaly detection defenses has an adverse effect as the clients with diverse enough data would also be removed. This presents a trade-off between fairness and robustness, which is also mentioned in \cite{kairouz2019advances} but unexplored in recent works. 

\paragraph{Data Sanitisation.} As a common technique to defend backdoor and poisoning attacks in FL, data sanitisation \cite{cretu2008casting} is employed in anomaly detection to filter out suspicious training examples. 
However, stronger data poisoning attacks are likely to break data sanitisation defenses \cite{koh2018stronger}.

\paragraph{Pruning.} Another defense technique against backdoor attacks is ``pruning''. Rather than directly filtering out the data, this technique evaluates if a unit is supposed to be inactive on clean data but activated in the updates \cite{liu2019lifelong}. However, the access to clean holdout data violates the privacy principle of FL, and is therefore the biggest concern of this defense technique.

\subsection{Gradient attack and solutions}

Although FL is designed to train a machine learning model without the access to clients' private data, recent research works have revealed that its default setting still suffers from privacy leakage attacks by gradient-based reconstruction \cite{geiping2020inverting,zhao2020iDLG,zhu2019deep}. 
Gradient attack, also known as client privacy leakage attack, is able to accurately reconstruct the private training data in the local client, given only a local gradient update shared from a client to central server.
So in gradient privacy attack, we assume that clients can be compromised in a limited manner. That is, an attacker cannot directly access the private training data $\mathcal{D}$ but only the gradient updates $\nabla\bold{w}_i$ calculated by SGD on $\mathcal{D}$. 
The gradient update data is intercepted by a malicious attacker via a compromised central server or eavesdropping of insecure network communication. 
This not only violates local data privacy but also the federated learning system by monitoring client confidential data illegally and silently, exposing federated learning into privacy leakage attacks. 

\paragraph{Client Privacy Leakage Attack.} 
This attack technique is to conduct a gradient-based feature reconstruction, where the adversary aims at developing a learning-based reconstruction algorithm that takes the gradient update $\nabla \bold{w}_i(t)$ at step $t$ to reconstruct the private data used to calculate the gradient. 
In the following, we mainly focus on the application scenarios of computer vision in federated learning, where the inputs are images or videos. Therefore, the learning-based reconstruction algorithm will begin from a random attack seed that can be a dummy image with the same resolution as that of the local client. Then, we can perform a forward inference given the dummy image and compute a gradient loss by measuring spatial distance between the current gradient w.r.t the model parameters and the actual gradient from the client. Then, we minimize the gradient loss w.r.t the dummy-initialized image to approach actual client private data. 
Thereby, the goal of this reconstruction algorithm is to iteratively modify the dummy image and finally approximate the original data in a local device if the gradient loss is converged to be minimal. 
In summary, this learning-based reconstruction algorithm fixes the parameters of a neural network model and tries to optimise the dummy-initialized image with regard to the gradient loss. Namely, compared to traditional machine learning, this algorithm takes the model $\bold{w}$ as its input but takes the image as its learnable parameters.

\begin{figure}[t]
    \centering
    \includegraphics[width=\linewidth]{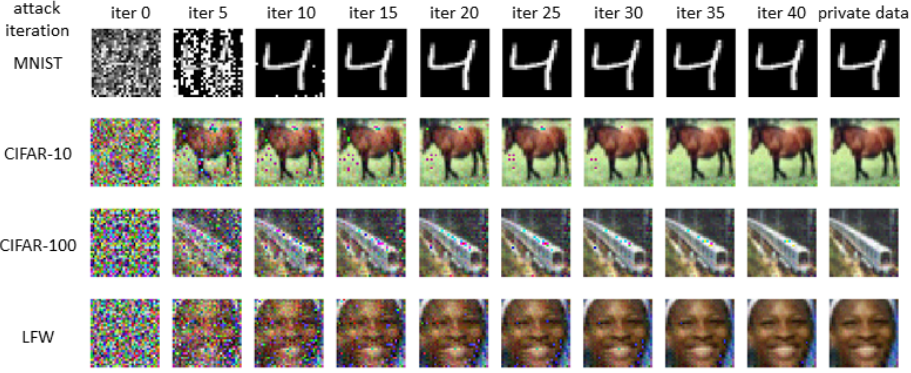}
    \caption{Examples of gradient attack, which are copied from \cite{wei2020gradient}. }
    \label{fig:ex_gradient}
\end{figure}

The open question still remains about how to effectively resist gradient-based client privacy leakage attack. The most effective way, of course, is to prevent the attack from its sources, e.g., encrypting network communication and strengthening server firewalls. When intercepting or eavesdropping is inevitable, differential privacy can be leveraged to eliminate privacy leakage.

\paragraph{Differential Privacy.}
As a widely-applied privacy-preserving technique, differential privacy is proposed to add noise into client privacy data, which thus prevents privacy leakage of personal data \cite{dwork2006differential}. Meanwhile, this comes with an acceptable cost of statistic data quality loss, which is caused by random noise from each client.
In this gradient-based attack scenario, differential privacy is implemented by adding noise to the gradient updates from the clients of the federated learning system, which thereby makes it more difficult to reconstruct the client data. 

However, this privacy-preserving technique introduces randomness into the gradient updates, and thus leads to model degeneration. This technique also makes the central server difficult to check the uploaded models from its clients, possibly resulting in conflicts with aforementioned anomaly detection. 

Fortunately, there are also other ways to perform defenses against gradient-based attacks by increasing attack difficulty, which do not compromise performance degeneration to the same extent. As suggested by \cite{wei2020gradient}, according to the principles of client privacy leakage attacks, we can lift the attack difficulty by either adding more parameters for an adversary or increasing estimation complexity. Therefore, this can be reached by (1) increasing batch size of mini-batch SGD in the clients, (2) lifting training image or video resolution in local clients, (3) making more steps of gradient updates before uploading to the central server, and (4) changing activation function in the local model.

\section{Data Heterogeneity Challenge for Federated Learning} 
Heterogeneity widely exists in healthcare and medical data. It is introduced by not only the variety of modalities, dimensionality and characteristics, but also the data acquisition differences. For example, the medical devices with different brands or local demographics can lead to significantly different source data distributions \cite{rieke2020future}. FL has addressed this heterogeneous data distribution issue as a critical challenge and a number of FL algorithms are proposed to tackle heterogeneity in federated networks \cite{yang2019federated}.

\subsection{Statistical Heterogeneity}

\begin{figure}[t]
    \centering
    \includegraphics[width=\linewidth]{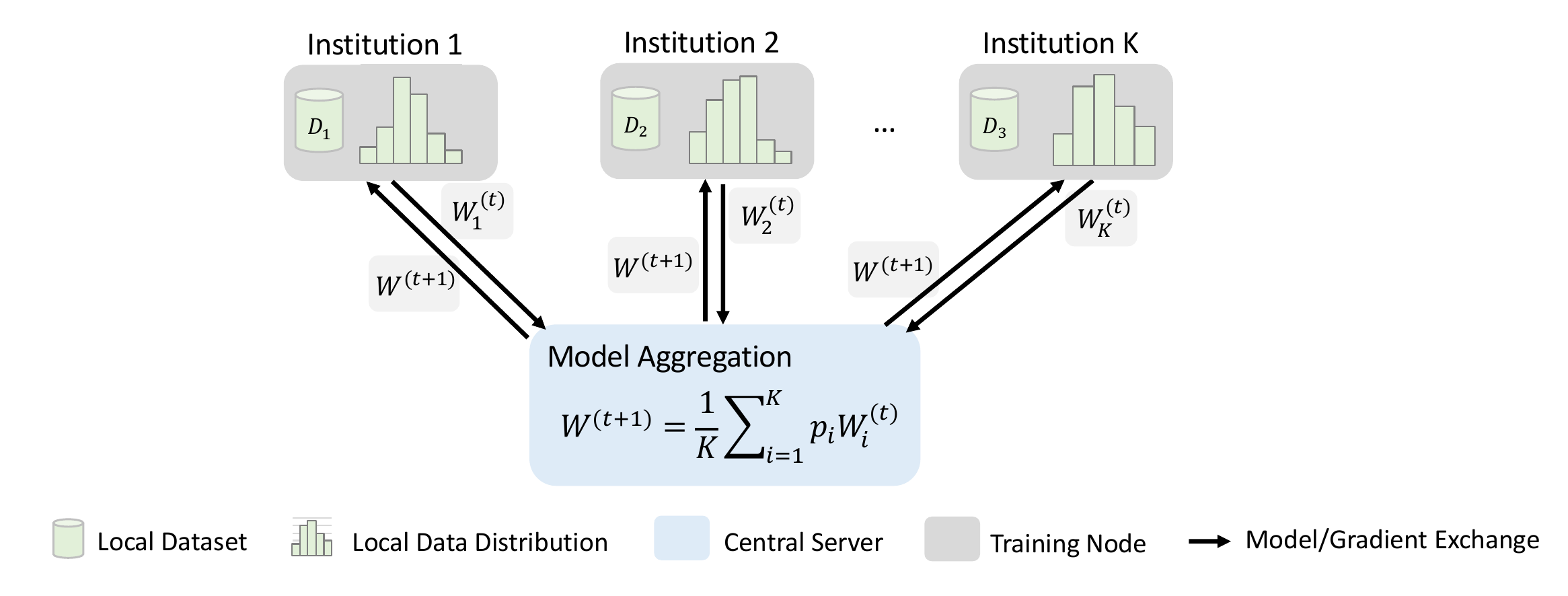}
    \caption{Statistical heterogeneity problem in federated learning.}
    \label{fig:heter_stat}
\end{figure}

Conventional machine learning is always built upon an independently and identically distributed (IID) assumption of a uniform dataset. However, statistical heterogeneity is an inherent characteristic across the distributed medical data providers, which can be identified as a non-IID problem, as is shown in Fig. \ref{fig:heter_stat} . Usually, medical data is not uniformly distributed across different institutions and sometimes can introduce a local bias \cite{li2019privacy,sheller2018multi}. Recently, some methods have been proposed to handle the statistical heterogeneity in the FL setting. These methods can be applied to most open health scenarios to enhance the robustness of existing distributed machine learning methods and tolerate heterogeneity across data providers. There are two main types of solutions for this, one is to cluster or group the data providers, and each cluster/group will have a unique global model, the other is by personalisation techniques where each medical data provider has both the shared part and the personalised part of model \cite{long2020federated}.

\paragraph{Clustered FL.}
According to the input data distribution, models with the same or similar data distribution can be clustered into the same cluster. Weight aggregation can be done inside the cluster. Below are the related methods for clustered federated learning.

In \cite{xie2020multi}, the authors propose a novel multi-center aggregation mechanism for heterogeneous federated learning which is common in various real-world applications. Multiple global models are learned from the non-IID data by optimising the predefined objective function, namely multi-center federated loss. In particular, the loss function of the proposed framework is defined as 
\begin{equation}
\sum_{i=1}^{n} p_{i} \cdot \min _{k}\left\|W_{i}-W^{(k)}\right\|^{2},
\end{equation}

where $p_{i}$ is the weight that is typically proportional to the size of the $i$-th client local dataset, $W_{i}$ is the parameter of the $i$-th client's local model and $W^{(k)}$ is the parameter of the $k$-th cluster. Any distance metric can be integrated into this framework. This paper takes the simplest L2 distance into consideration and uses K-means as the clustering method.

The distance measurement for clustering methods can be of different forms. Apart from L2-distance, there are more sophisticated distance measurements represented in a hierarchical form, which can be leveraged for specific sets of medical data silos. In \cite{sattler2020clustered}, geometric properties of the FL loss surface is used for client clustering. Clients within the same cluster have jointly trainable data distributions. 

Although clustered FL better utilizes the similarity among the users, there are still challenges when performing it in open health. For example, the cluster identities of a hospital or clinic are usually unknown, so it is essential to identify the cluster membership of these data providers first and then optimise each of the cluster models in a distributed setting. To achieve this efficiently, \cite{ghosh2020efficient} designs a framework known as Iterative Federated Clustering Algorithm (IFCA) for clustered FL. IFCA alternately estimates the cluster identities and minimises the loss functions so as to allow the model to converge at an exponential rate with a relaxed initialisation requirement.

\paragraph{Personalised FL.}

Personalised FL aims to provide personalised services to participating medical institutes or patients based on the their medical images, fragmented data sources and healthcare data with privacy concerns.

The balance between individual learning and collaboration has been discussed in \cite{deng2020adaptive} in a theoretical way. As a result, when the user’s data distribution does not deviate too much from the data distribution of other users, collaboration can be beneficial to reduce the local generalisation error. When some users data is too different from the data of others to represent the overall data distribution, independent training is preferable.

Sometimes, it is possible to train one personalised model per client. A theoretic study of personalisation in FL is presented in \cite{mansour2020three}. The authors propose to use data interpolation as a personalisation technique.

Data interpolation realises personalisation by domain adaptation. Local dataset $D_k$ is regarded as target domain, and global dataset or the cluster dataset $\mathcal{C}$ is regarded as source distribution. The objective function is optimised based on concatenated dataset,
\begin{equation}
\lambda \cdot \mathcal{D}_{k}+(1-\lambda) \cdot \mathcal{C},
\end{equation}
where $\lambda$ is a hyperparameter that can be obtained by cross validation.

Some other popular methods for personalised FL can be categorized as personalisation layers where part of the layers are shared and aggregated across multiple clients. The rule of shared layer selection can vary. For example, in \cite{arivazhagan2019federated}, representation layers act as the personalised components, while the decision layers are shared across participants. By contrast, in \cite{liang2020think}, the representation layers are shared across participants, and decision layers remain local as a personalised component.

\subsection{Model Architecture Heterogeneity}

\begin{figure}[t]
    \centering
    \includegraphics[width=\linewidth]{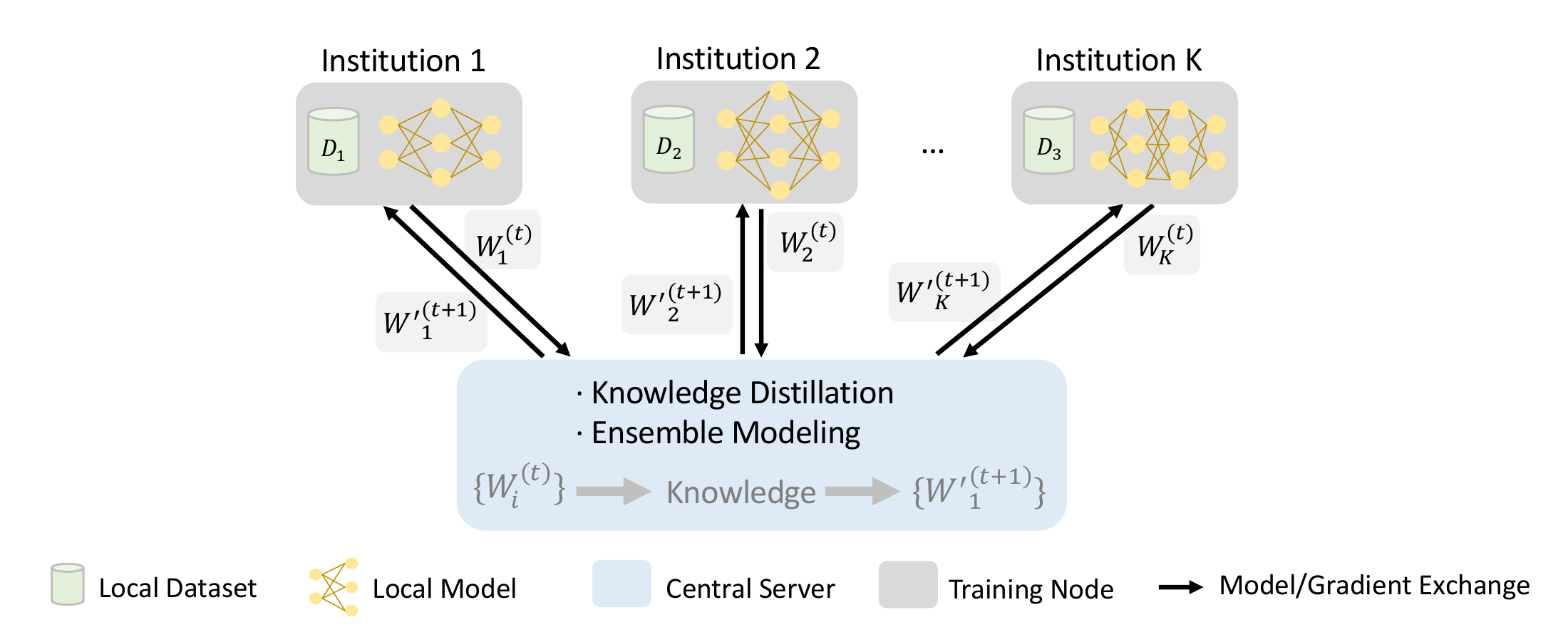}
    \caption{Model heterogeneity problem in federated learning.}
    \label{fig:heter_model}
\end{figure}

Medical data providers tend to use robust statistical models from their local medical data, which is collected in huge amounts by modern healthcare systems \cite{rieke2020future}. This brings about model heterogeneity to an FL system. Model architecture heterogeneity in Fig. \ref{fig:heter_model}, as the main form of model heterogeneity, will hinder the model aggregation procedure in traditional FL algorithms.

To solve this problem and train a robust statistical model from distributed medical data, knowledge distillation, proposed by Hinton et al. \cite{hinton2015distilling} in 2015, is widely used to transfer knowledge between models with different architectures. \cite{lin2020ensemble} proposes ensemble distillation for robust model fusion. It allows for heterogeneous client data and models, and clients with different neural architectures to be considered in one FL system. It defines $p$ distinct model prototypes with different architectures. During the aggregation procedure in each communication round, all received client models are distilled to $p$ model prototypes without additional computation burden on clients. Then, the fused prototype models are sent to activated clients for the next round.

The authors in \cite{tan2021fedproto} also borrow the concept of prototypes and use them to represent classes rather than models. Prototype aggregation is applied in the server to solve the heterogeneous setting in FL.

Sometimes, as a result of various dataset sizes and computation abilities, the model for a specific medical data provider is independently designed. This makes the model unique, and the use of traditional averaging aggregation no longer possible. To solve this problem, Li and Wang propose Federated Learning via Model Distillation (FedMD) which is a universal framework enabling FL to work with uniquely designed local models \cite{li2019fedmd}. Assume there are $k$ clients, each client owns not only a private dataset $\mathcal{D}_{k}$ but also a public dataset $\mathcal{D}_{0}$, $k=1 \ldots m$. Each client computes the class scores $f_{k}\left(x_{i}^{0}\right)$ on the public dataset, and transmits the result to a central server. Instead of model parameter aggregation, the server computes an updated consensus on the average of these class scores. This is realised through knowledge distillation that can transmit learned information in a model-agnostic way.

Although knowledge distillation works well for the model architecture heterogeneity challenge, the communication and computation costs remain a problem. Jeong et al. propose to minimize the communication overhead while enjoying the benefit brought by a massive amount of private data providers \cite{jeong2018communication}. A new distributed training algorithm named federated distillation (FD) has been developed to improve communication efficiency for an on-device machine learning framework. It exchanges the model output rather than the model parameters, which significantly decreases the communication payload from model size to output size.

\subsection{Transfer learning for cross-silo federated learning}

Existing medical data is not fully exploited by traditional machine learning methods because it sits in data silos and privacy concerns restrict access to this data. As a result of insufficient data, a gap between research and clinical practice exists. FL provides an opportunity to take full advantage of cross-silo data and significantly contribute to open health in future decades.

Transfer learning, as a special case of machine learning, is introduced to solve the heterogeneity problem, expand the scale of medical datasets and further improve the performance of the final model \cite{liu2018secure}. For the distributed medical data silos with different kinds of private patient data, the combination of transfer learning and federated learning can lead to a flexible framework adapted to various secure multi-party machine learning tasks \cite{xu2019federated}.

Classical FL methods like FedAvg require that all contributors share the same feature space \cite{mcmahan2017communication}. However, the scenario with such common entities is rare in reality. In most cases, data contributors share heterogeneous feature spaces and/or model architectures. The authors in \cite{liu2018secure} address this limitation of existing FL approaches and utilize transfer learning to provide solutions for the entire feature space under a federation. Their work has formalised the federated transfer learning (FTL) problem in a privacy-preserving setting where all parties are honest-but-curious. Next, they propose an end-to-end method to solve the FTL problem. Compared with non-privacy-preserving transfer learning, the proposed method achieves comparable performance in terms of convergence and accuracy. Moreover, novel privacy preserving techniques, i.e. homomorphic encryption (HE) and secret sharing can be incorporated with learning models, i.e. neural networks, under the proposed FTL framework without much modifications.

Such privacy-preserving FTL solution is well suited for cross-silo open health scenarios, because it is a well-developed framework that takes all the related aspects into account, including performance, scalability, computation and communication. FTL is superior to non-federated self-learning approaches, and performs as well as non-privacy-preserving approaches.

\section{Discussion}
\label{sec::discussion}

In the above sections, we identified key challenges and possible solutions for FL in the healthcare industry. Here, we discuss the implications of solving the healthcare challenges and explore the benefits of a successful open innovation framework with FL.

\subsection{Implications of solving the security, privacy and heterogeneity challenges}

If the existing security, privacy and heterogeneity challenges can be solved, there are a number of implications for various industries, including healthcare. Below we detail three primary implications.

\paragraph{Shared knowledge and expertise.}
Having a secure way to learn from various local health data sets will increase shared knowledge and expertise among different healthcare institutions. Organisations is likely to be more willing to use their data for collaborative research when they can ensure the privacy of their own patient data and are not required to provide copies of their data. Certainly, it will be essential to identify and engage organisations and institutions that are willing to explore the use of FL for healthcare purposes and to have legal and privacy experts who can verify whether methods comply with existing privacy standards and regulations. Solving the heterogeneity problem will also drive knowledge sharing as it will allow for diverse data to be used to develop more useful global models.

\paragraph{Adoption of deep learning.}
Deep learning has already demonstrated potential to enhance patient care in the healthcare industry. Previous reviews \cite{reviewSolares2020,reviewShickel2017,reviewXiao2018} present numerous applications of deep learning to Electronic Health Records (EHRs) and deep learning has also shown success with image data from medical images and tissue samples \cite{Ker2018}. Despite the potential for deep learning to improve patient care, adoption in the healthcare industry is slow. This is often due to models being developed in single healthcare institutions using single datasets, resulting in a lack of robustness across populations \cite{xu2019federated}.

If healthcare information can be shared with the help of an FL system, this will overcome existing limitations. Healthcare institutions that previously didn't adopt deep learning due to potential bias issues that can arise from training in single institutions, would now have access to models with better generalisability due to training on larger and more representative populations. Furthermore, smaller institutions which were incapable of building predictive algorithms due to small data sizes, will now have access to risk prediction models. This will provide these institutions with increased functionality that can be used to help improve clinical care.

\paragraph{Generalised methodologies.}
The security and privacy challenges, as well as the data heterogeneity problems mentioned above, are not only found in health data. Therefore, designing methods that can address these challenges will facilitate the healthcare sector but will also be relevant to other industries with sensitive data (such as banking or insurance). Methodologies developed could be directly applied to these other relevant industries. Furthermore, general FL research will help advance the field of FL, and in a time where learning joint models from siloed datasets can provide immense potential, particularly in healthcare, it is paramount that the development of FL algorithms be continued.

\subsection{Benefits of an open innovation framework with FL to healthcare}

If a successful open innovation FL system was implemented in the healthcare sector, it stands to benefit in a multitude of ways. These benefits will be received by participants, patients and the healthcare system.

\paragraph{Participants.}
In addition to the benefit of knowledge and expertise sharing as indicated in the above section, participants also stand to benefit from FL system through the development of collaborative partnerships. These types of frameworks are the very definition of open innovation, as various external parties will be involved to provide expertise, including clinicians, healthcare professionals, healthcare institution managers, data scientists and software engineers. 
We expect that implementation of FL systems will lead to long-term partnerships between medical centres, hospitals, and governments.

\paragraph{Patients.}
Because FL systems will enhance the adoption of deep learning algorithms, implementation of these algorithm has potential to directly benefit patients and their outcomes in both the cross-device and cross-silo settings.

In the cross-device setting, patients can benefit through shared information from wearable devices. Wearable devices can be used for monitoring health and safety of patients, managing chronic diseases, assisting in the diagnosis and treatment of conditions, and for monitoring rehabilitation \cite{Lu2020wearables}. For example, in \cite{Jung2015wearables}, data from wearable devices was used to create a fall detection system that could produce emergency alerts so that immediate treatment could be provided to patients. Wearables have also been used by patients with chronic obstructive pulmonary disease to help screen for early disease deterioration \cite{Wu2018wearables}. They have also found relevance in rehabilitation, helping to understand stroke recovery and modify treatment plans in line with recovery progress \cite{Jayaraman2018wearables}. With the help of FL, model information from individual patient devices can be used to help develop more generalised models that offer improvements in identifying falls, disease deterioration and patient monitoring. This would help optimise care for patients and lead to more personalised treatments, driving improvements in patient outcomes.

In the cross-silo setting, patients can benefit through shared information across healthcare institutions. Much research to date on machine learning and deep learning for healthcare data has utilised the Medical Information Mart for Intensive Care (MIMIC-III), due to it being a freely-available database. It has been used for the development of deep learning algorithms to predict mortality, sepsis, future diagnosis and hospital readmissions \cite{Hou2020mimic, Scherpf2019mimic, Lin2018mimic, Ma2018mimic}. All these outcomes have potential to prevent poor health outcomes for patients and provide optimised care at end-of-life. However, MIMIC-III data is from a single healthcare centre and includes only patients that are admitted to the Intensive Care Unit (ICU) in the hospital. Therefore many of the predictive models developed on this dataset will not perform equivalently in other health datasets. By using FL, global models can be developed from multiple healthcare institutions that offer improved performance due to training on larger and more diverse health data \cite{xu2019federated}. These improved predictive models can provide additional information to clinicians about risks and benefits of different treatment options and outcomes \cite{xu2019federated}. This has potential for more effective treatment earlier, leading to improved patient outcomes.

The promise of FL to improve patient outcomes has already been demonstrated in the healthcare sector, for both risk prediction and identifying similar patients. \cite{Brisimi2018} used an FL framework to develop a prediction model for hospitalisations due to heart diseases based on information in EHRs. The decentralised model achieved similar performance to the centralised method and the authors extracted important features to facilitate interpretability. \cite{Sharma2019} compared an FL framework with a centralised approach for predicting in-hospital mortality and also found that the FL approach provided comparable performance to the centralised setting. \cite{Lee2018} used a privacy-preserving federated environment to identify similar patients across healthcare institutions without sharing patient-level information. Similarly, \cite{kim2017federated} performed computational phenotyping without sharing patient-level data. These examples demonstrate the success of FL compared to typically centralised approaches and its potential in the healthcare industry to improve care of patients.

\paragraph{Healthcare system.} Given the current pressure for healthcare systems to enhance sustainability, an FL system is an attractive option. With the use of FL, it is possible to develop more generalisable models that can assist in providing clinical care. More effective and targeted treatment of patients may result in less time spent visiting emergency departments and hospitals, slowed disease progression for chronic disease, and better outcomes sooner \cite{rieke2020future}. This has potential to reduce healthcare system costs, hence contributing to a more sustainable healthcare system.

\paragraph{Data Structure.} The healthcare activity may produce data with various structure, for example, images from medical imaging, texts on clinic reports, tabular data in hospital's database, sequential patient journey in healthcare service system, and time series from wearable devices or ICU. Different data structure needs to tackled with different data processing technique. Moreover, the data fusion of multi-modal data is also a common solution to build intelligent healthcare application. To model the complex data of real-world is a practical challenge. This chapter demonstrates the FL framework using image data. However, the discussed problem and solution can be generally applied to a broad scenario with different data structures.

\section{Conclusion}
\label{sec::conclusion}

FL holds promising potential to enable shared healthcare information, knowledge and expertise between institutions, while preserving privacy of individuals. Although there remain challenges for data security, privacy and heterogeneity, this is an active area of FL research with solutions already being identified. The implications of FL for health are many, including facilitating sharing of healthcare information, increasing adoption of deep learning algorithms that can produce more generalised models, and the development of improved methodologies that are applicable to industries beyond healthcare. Long-term benefits will result from FL-enabled open innovation of health, which will be felt at the participant, patient and healthcare system level. 

We believe that FL will help leverage existing health data assets to directly impact patient care and therefore, offers immense opportunity for open innovation in the healthcare sector.

\bibliographystyle{plain}
\bibliography{references}

\end{document}